\documentclass[twocolumn,showpacs,preprintnumbers,prb,amsmath,amssymb]{revtex4}

\usepackage[dvips]{graphicx}
\usepackage{dcolumn}
\usepackage{bm}
\usepackage{amsmath}

\begin{document}

\title{Mesoscopic Fano effect in a Quantum Dot Embedded in an
Aharonov-Bohm ring}

\author{Kensuke Kobayashi, Hisashi Aikawa} 
\author{Shingo Katsumoto}
\author{Yasuhiro Iye}

\affiliation{ Institute for Solid State Physics, University of Tokyo,
5-1-5 Kashiwanoha, Chiba 277-8581, Japan}


\begin{abstract}
The Fano effect, which occurs through the quantum-mechanical
cooperation between resonance and interference, can be observed in
electron transport through a hybrid system of a quantum dot and an
Aharonov-Bohm ring.  While a clear correlation appears between the
height of the Coulomb peak and the real asymmetric parameter $q$ for
the corresponding Fano line shape, we need to introduce a complex $q$
to describe the variation of the line shape by the magnetic and
electrostatic fields.  The present analysis demonstrates that the Fano
effect with complex asymmetric parameters provides a good probe to
detect a quantum-mechanical phase of traversing electrons.
\end{abstract}


\maketitle

\section{Introduction}
\label{SecIntro}
An Aharonov-Bohm (AB) ring and a quantum dot (QD) have been typical
mesoscopic systems that continually invoke fundamental interest in the
researchers. The wave nature of electron manifests itself in the
former while the particle nature of electron features in the latter.
Combination of these complementary systems into a hybrid one enables
us to explore the problem how coherent is the transport of electrons
through a QD, where many electrons interact with each other.  In 1995,
Yacoby \textit{et al.}~\cite{YacobyPRL1995} performed a pioneering
work to tackle this issue by using such a system, namely, a QD
embedded in one arm of an AB ring. It was found that an electron at
least partially maintains its coherence in passing the
QD.\cite{YacobyPRL1995,kats1996} As regards the phase of the AB
oscillation, although phase lapse was found across the Coulomb peak,
it was subsequently recognized to be due to the two-terminal nature of
the device,\cite{YeyatiPRB1995,YacobyPRB1996} thus revealing that
intrinsic phase measurement is a nontrivial issue.  To avoid the
problem, four-terminal measurements were performed and it was found
that each level inside the QD acts as a Breit-Wigner-type scatterer in
that the phase of electron smoothly changes $\pi$ at the
resonance.\cite{SchusterNature1997} At the same time, however, it has
been found contrary to a naive expectation that the peaks in the
Coulomb oscillations are in phase with each other and an unexpected
phase lapse occurs at the middle of the Coulomb valley.  Such an
interferometry has also been applied to QD's in the Kondo
regime,\cite{vanderWielScience2000,JiScience2000,JiPRL2002} while the
results have not converged, again revealing the
difficulty.\cite{EntinPRL2002}

In the above experiments, the transport properties of a QD have been
the main scope and the other arm with no QD has served as
``reference.''  However, one might ask what happens if the coherence
of the QD and the AB ring is fully maintained and thus the arm as well
as the QD should be equally treated. In this situation, Fano effect
has been expected to
occur,\cite{DeoMPL1996,RyuPRB1998,KangPRB1999,WohlmanCM2001,
HofstetterPRL2001,KimPRB2002,UedaJPSJ2003} and was experimentally
established in our previous study.\cite{KobayashiPRL2002}  

\begin{figure}[htb]
\center 
\includegraphics[width=0.8\linewidth]{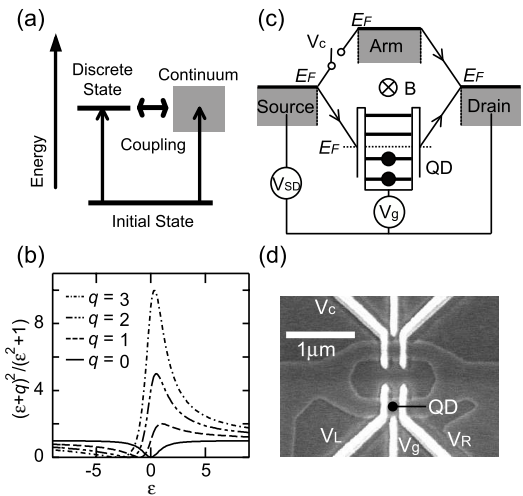}
\caption{(a) Principle of the Fano effect.  (b) Fano's line shapes for
several real $q$ parameters. (c) Schematic representation of the
experimental setup (see text).  (d) Scanning electron micrograph of
the correspondent device.}
\label{SampleFig} 
\end{figure}

Let us briefly summarize Fano theory.\cite{FanoPR1961} Consider a
system with a discrete energy state embedded in the continuum, which
can be described by the following Hamiltonian [see
Fig.~\ref{SampleFig}(a)]:
\begin{eqnarray}
{\cal H} &=& E_{\varphi}|\varphi\rangle \langle \varphi| +
\sum_{E'}E'|\psi_{E'}\rangle \langle \psi_{E'}|\nonumber\\
&+& \sum_{E'} \left(
V_{E'}|\psi_{E'}\rangle\langle \varphi|+
V_{E'}^*|\varphi\rangle\langle \psi_{E'}| \right),
\label{FanoEq1}
\end{eqnarray}
where $\varphi$ is a discrete state with energy $E_{\varphi}$ and
$\psi_{E'}$ is a state in the continuum with energy $E'$.  $V_{E'}$ is
the coupling strength between the discrete state and the continuum.  A
system eigenstate $\Psi_{E}$ can be expressed by a linear combination
of $\varphi$ and $\{\psi_{E'}\}$ with the coefficients analytically
obtained.  We consider a situation that an incoming initial state $i$
interacts with the system through a perturbation $\cal T$ and ends up
as a linear combination of $\{\Psi_E\}$. The evolution from $i$ to
$\Psi_E$ consists of two paths: one is through continuum states and
the other is through a ``resonant state" $\Phi$, which is a
modification of $\varphi$ through the interaction $V_{E'}$ with the
continuum.  The transition probability has a peculiar property around
the resonance energy as
\begin{equation}
\frac{|\langle \Psi_{E}|{\cal T}|i\rangle|^2}{|\langle \psi_{E}|{\cal
T}|i\rangle|^2}= \frac{(\epsilon +q)^2}{1+\epsilon^2}
\quad\mbox{with } \epsilon=\frac{E-E_{0}}{{\it \Gamma}/2},
\label{FanoEqn21}
\end{equation}
where $E_0$ and ${\it \Gamma}$ are the energy position and width of the
resonance state, respectively.  The so-called ``Fano's asymmetric
parameter" $q$ is defined as
\begin{equation}
q \equiv \frac{\langle \Phi |{\cal T}|i\rangle}{\pi V_{E}^{*}\langle
\psi_{E}|{\cal T}|i \rangle},
\label{defq}
\end{equation}
which is a measure of the coupling strength between the continuum
state and the resonance state (namely, the strength of the
configuration interaction).  Several curves for different $q$'s are
plotted in Fig.~\ref{SampleFig}(b).

The Fano effect is an expansion of the Breit-Wigner-type resonance
scattering and was established in the atomic physics more than 40
years ago. Since then, reflecting the generality of the original model
of Eq.~(\ref{FanoEq1}), the Fano effect has been found ubiquitous in a
large variety of experiments including neutron
scattering,\cite{AdairPR1949} atomic
photoionization,\cite{FanoRau1986} Raman
scattering,\cite{CerdeiraPRB1973} and optical
absorption.\cite{FaistNature1997}  As this effect is essentially a
single-impurity problem, an observation on a single site would reveal
this phenomenon in a more transparent way.  The most sensitive probe
for such single-site experiments is the electron transport.  Compared
to its long history in the spectroscopy, however, the general
importance of Fano effect in the transport was only recently pointed
out in Refs.~\onlinecite{TekmanPRB1993} and
\onlinecite{NockelPRB1994}, although its characteristic feature has
already appeared in the theoretical studies.\cite{ButtikerPRB1984}
Experimentally, the single-site Fano effect has been reported in the
scanning tunneling spectroscopy of an atom on the
surface~\cite{MadhavanScience1998,LiPRL1998} and in transport through
a QD.\cite{GoresPRB2000,ZachariaPRB2001}  While the latter case is
the first observation of this effect in a mesoscopic system, there is
no well-defined continuum energy state and the mechanism for the
appearance of the effect remains as an intriguing puzzle.

As mentioned, the QD-AB-ring hybrid system with parameters
appropriately tuned provides a unique Fano system where the transition
process shown in Fig.~\ref{SampleFig}(a) is realized in real space as
schematically sketched in Fig.~\ref{SampleFig}(c).  An electron
traverses from the source to the drain along two interfering paths.
One is through the discrete state in the QD and the other is through the
arm.  In the previous paper~\cite{KobayashiPRL2002} we reported the
following: (i) The electronic states in the QD-AB-ring system can be
sufficiently coherent for the Fano effect to emerge in transport.
(ii) The coherent Fano state disappears at finite source-drain
voltage.  (iii) The phase difference between the two paths can be
controlled by magnetic field, resulting in characteristic variation in
the line shape of the resonant peaks, which suggests the necessity to
treat the parameter $q$ as a complex number.

Regarding this issue, many problems are still open. For instance, the
phase controlling of the Fano effect should be clarified in terms of a
complex $q$. In addition, we would list the phase evolution of the
electrons through a QD in the presence of the Fano
effect,\cite{AharonyPRL2003,KimPRB2003} the Fano effect as a probe of
phase coherence,\cite{ClerkPRL2001} the Fano effect in the Kondo
regime,\cite{HofstetterPRL2001} and the detailed mechanism for the
zero-bias resonance peak/dip in the differential conductance through a
Fano system.\cite{GoresPRB2000,KobayashiPRL2002}

In this paper, after briefing the experimental setup in
Sec.~\ref{SecExperiment}, we extend the previous work to quantitative
analysis of the experiments in Sec.~\ref{SecFanoResult}.  First, we
report the application of the conventional Fano formula with a real
$q$ to our observation, and discuss the relation between the original
Coulomb peaks and the Fano peaks in Sec.~\ref{subsecA}.  Then, we
focus on the magnetic and electrostatic phase controlling of the Fano
effect in Secs.~\ref{subsecB} and \ref{subsecC}, respectively. This
subject is the central part of this paper, where we show that Fano
effect can be a powerful tool to investigate the electron phase
variation in such mesoscopic transport.  In Sec.~\ref{subsecB}, we
actually derive complex $q$ as a function of the magnetic field from
the experiment and compare it with the theory. The electrostatic
controlling of the Fano line shape is similarly analyzed in
Sec.~\ref{subsecC}.  We also discuss the phase evolution over multiple
resonances and the temperature dependence of the Fano effect in
Secs.~\ref{subsecD} and \ref{subsecE}, respectively.

\section{Experiment}
\label{SecExperiment}
Figure~\ref{SampleFig}(d) shows a scanning electron micrograph of the
device fabricated by wet-etching a two-dimensional electron gas (2DEG)
in an AlGaAs/GaAs heterostructure (mobility $=9\times 10^5$~cm$^2/$V~s
and sheet carrier density $=3.8\times10^{11}$~cm$^{-2}$).  The length
of one arm of the ring $L$ is $\sim 2$~$\mu$m.  Two sets of three
fingers are Au/Ti metallic gates to control the local electrostatic
potentials of the device.  The three gates ($V_R$, $V_L$, and $V_g$)
at the lower arm are used for controlling the parameters of QD (with
area about 0.15$\times$0.15~$\mu$m$^2$) and the gate at the upper arm
is to apply $V_C$, which determines the conductance of the upper arm.
Measurements were performed in a mixing chamber of a dilution
refrigerator between 30~mK and 800~mK by a standard lock-in technique
in the two-terminal setup with an excitation voltage of 10~$\mu$V
(80~Hz, 5~fW) between the source and the drain.  Noise filters were
inserted into every lead below $1$~K as well as at room temperature.

As sketched in Fig.~\ref{SampleFig}(c), this single-site Fano system
is tunable in several ways.  In the addition energy spectrum, the
discrete energy levels inside the QD are separated by respective
quantum confinement energy and the single-electron charging energy
$E_c$, but we can shift the spectrum by the center gate voltage $V_g$
to tune any one of them to the Fermi level.  The coupling between the
continuum and the levels in the QD is controlled by $V_R$ and $V_L$.
$V_C$ can control the conductance of the interference path as well as
the phase shift of the electrons traversing underneath it (details are
given later).  The controlling of the phase difference between the two
paths ($\Delta \theta$) is also possible by the magnetic field $B$
piercing the ring.

\begin{figure}[htb]
\center \includegraphics[width=0.85\linewidth]{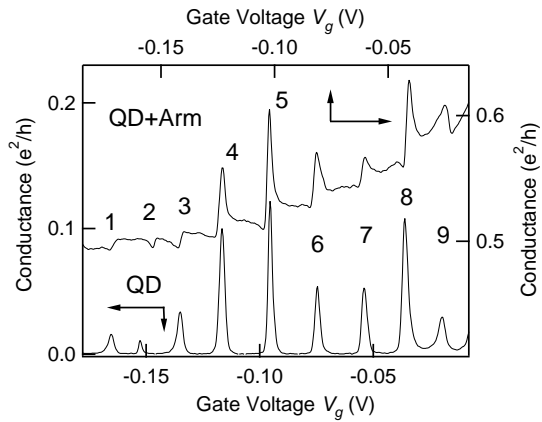}
\caption{Typical Coulomb oscillation at $V_C = -0.11$~V with the arm
pinched off, and asymmetric Coulomb oscillation at $V_C = -0.085$~V
with the arm transmissible. The latter shows a clear Fano effect.  The
Coulomb and Fano peaks are in good one-to-one correspondence as they
were numbered consecutively from 1 to 9. Both of them were obtained at
$T=30$~mK and $B=0.91$~T.}
\label{TypicalFano}
\end{figure}

An advantage of the present sample structure (wrap-gate-type dot
definition) is clearly shown in Fig.~\ref{TypicalFano}.  The Coulomb
oscillations through the QD with and without the conduction through
the other arm show a clear one-to-one correspondence of the
conductance peaks. This is due to the weakness in the electrostatic
coupling between the two sets of gates ($V_C$ and \{$V_L$, $V_g$,
$V_R$\}), {\it i.e.}, the circuit is already defined by etching and
the gate electrodes can be very thin.  Still, though, we need a small
shift by 0.0065~V between the upper and lower axes in
Fig.~\ref{TypicalFano}, due to the electrostatic coupling between the
gates.

\section{Results and Discussion}
\label{SecFanoResult}
\subsection{Fano effect in the Coulomb oscillation}
\label{subsecA}
In Fig.~\ref{TypicalFano}, the peaks were numbered consecutively from
1 to 9. The small irregularity of the peak positions reflects the
variation of single-electron energy-level spacing in the QD and
indicates that the transport is through each single level.  This is
also consistent with the fact that the peak heights vary randomly.  In
the upper panel, the Fano effect emerges and the line shapes of the
peaks become very asymmetric and even change to a dip structure for
the peaks 1, 2, and 3.

We have found that the line shapes of the peaks in the upper curve
in the conductance $G_{tot}(V_g)$
can be well fitted to the following equation
\begin{equation} 
G_{tot}(V_g) = G_{bg} + G_{Fano}(V_g),
\label{CondFanoEqn}
\end{equation}
where $G_{bg}$ is noninterfering contribution of the parallel arm and
is a smooth function of $V_g$ that can be treated as a constant for
each peak.  $G_{Fano}$ is the Fano contribution expressed as
\begin{equation} 
G_{Fano}(V_g) = A \frac{(\epsilon +q)^2}{1+\epsilon^2} \quad
\mbox{with } \epsilon=\frac{\alpha(V_g-V_{0})}{{\it \Gamma}/2}.
\label{CondFanoEqn2}
\end{equation}
Here, $A$, $V_0$, and ${\it \Gamma}$ are the amplitude, the position,
and the width of the Fano resonance, respectively. $\alpha$ is the
proportionality factor which relates the gate voltage $V_g$ to the
electrochemical potential of the QD. In our system, $\alpha \sim
20$~$\mu$eV$/$mV as estimated from the differential conductance at the
finite source-drain voltages $V_{sd}$.\cite{KobayashiPRL2002}

\begin{figure}[htb]
\center \includegraphics[width=0.85\linewidth]{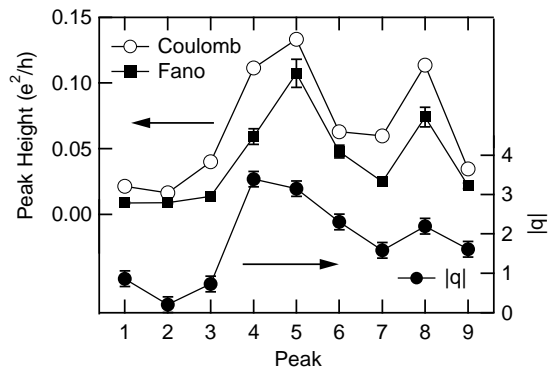}
\caption{Fitted $q$ values for the peaks 1--9 in
Fig.~\ref{TypicalFano} are shown in the lower panel. The heights of
Coulomb peaks and Fano peaks are also plotted in the upper panel. The
three of them are well correlated.}
\label{FanoCoulomb} 
\end{figure}

The fitted values ${\it \Gamma}$ are 40--60~$\mu$eV, reflecting the
variation of the coupling strength between the leads and the QD in the
original Coulomb peaks.  This ${\it \Gamma}$ is consistent with the
width of the zero-bias peak observed in the differential conductance
at a finite $V_{sd}$.\cite{KobayashiPRL2002}  Figure~\ref{FanoCoulomb}
shows the fitted $|q|$ for the nine peaks.  The signs of $q$ are
positive for all the peaks except the peak 9. This corresponds to the
observation that all the Fano features (the dips 1--3 and peaks 4--8)
except 9 have the asymmetric tails to the same direction.  It is
important that the signs of $q$ are same for many consecutive Fano
peaks since this directly reflects that the Coulomb peaks are in phase
as introduced in Sec.~\ref{SecIntro}. We will discuss it in
Sec.~\ref{subsecD}.

In the upper panel of Fig.~\ref{FanoCoulomb}, we plot the heights of
the Coulomb peaks and the Fano peaks. The height of a Fano peak is
defined as $A (1+q^2)$ in the fitting form of Eq.~\eqref{CondFanoEqn2},
which is the difference of the conductance maximum and minimum of the
Fano resonance. Not only are the heights of the Coulomb and Fano peaks
correlated but the observed $|q|$ values are also related to the peak
heights in a way that a large peak tends to have a large $|q|$.  This
can be qualitatively explained as follows. When the arm is pinched
off, electrons never pass the continuum, resulting in $|\langle
\psi_{E}|{\cal T}|i \rangle| \rightarrow 0$. This corresponds to $|q|
\rightarrow \infty$ in Eq.~(\ref{defq}) and
Eq.~\eqref{FanoEqn21} asymptotically gives a Lorentzian Coulomb peak as
seen in Fig.~\ref{SampleFig}(b).
This reflects the fact that the Fano theory includes the
Breit-Wigner scattering problem as a special limit of $|q|
\rightarrow \infty$, where there is no
path through the continuum, namely, there is no configuration
interaction.
The height of the Coulomb peak is proportional
to $|\langle \Phi |{\cal T}|i\rangle |^2$, which corresponds to the
coupling strength between the QD and the leads. Opening of the arm
means that $|\langle \psi_{E}|{\cal T}|i \rangle |$ is increased from
0 to a finite value with $|\langle \Phi |{\cal T}|i\rangle |$ fixed.
The asymmetric parameter $|q|$, therefore, is larger for a larger
original peak.

\begin{figure}[htb]
\center \includegraphics[width=0.85\linewidth]{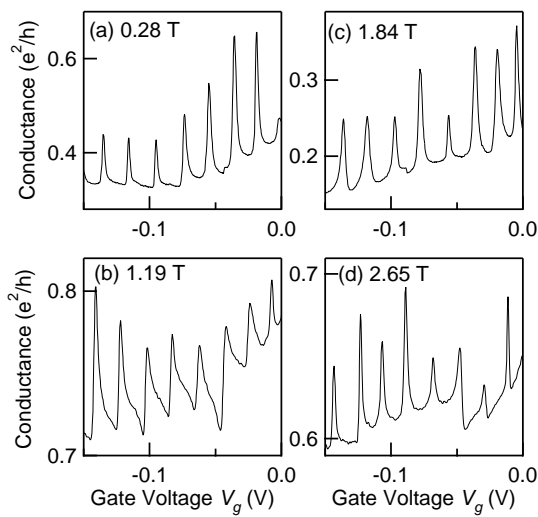}
\caption{\label{FanoMag} Fano's line shapes taken at (a) 0.28~T, (b)
1.19~T, (c) 1.84~T, and (d) 2.65~T. They were obtained at $T=30$~mK.}
\end{figure}

The above Fano effect was observed at $B \sim 0.91$~T.  We found the
Fano effect prominent within several specific magnetic-field ranges as
shown in Figs.~\ref{FanoMag}(a)-(d) besides $B \sim 0.91$~T, while it
was less pronounced in other ranges.  This implies that the coherence
of the transport through the QD strongly depends on $B$.  A similar
role of $B$ is reported in the Kondo effect in a
QD,\cite{vanderWielScience2000} where the Kondo effect appeared at
the specific magnetic field other than at $B \sim 0$~T. The phenomena
arise from the change of the electronic states caused by $B$ in a QD.
It is also possible that the magnetic field affects the coherent
transport in the AB ring, due to the reduction of boundary roughness
scattering by the magnetic field as theoretically treated in
Ref.~\onlinecite{AkeraPRB1991}.  An example of the experimental
indication may be seen in Ref.~\onlinecite{KouwenhovenPRB1989}.
Henceforth, we will focus on the Fano effect observed at $B \sim
0.91$~T.

\subsection{Magnetic-field control of the Fano interference}
\label{subsecB}
One of the experimental advantages of the present system over the
other canonical Fano systems with microscopic sizes lies in the
spatial separation between the discrete level and the continuum, which
allows us to control Fano interference both magnetically and
electrostatically.

\begin{figure}[htb]
\center
\includegraphics[width=0.85\linewidth]{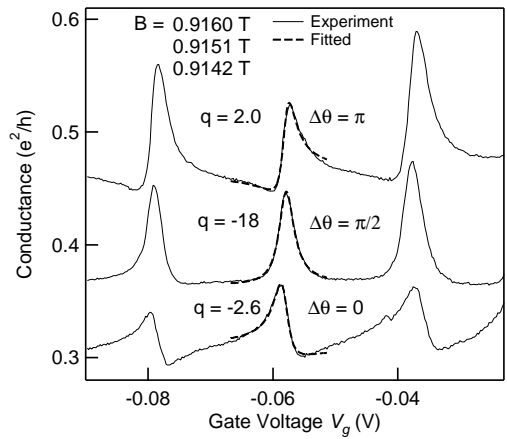}
\caption{Conductance of three Fano peaks at 30~mK at the selected
magnetic fields at $B=0.9142$, 0.9151, and 0.9160~T. They are
incrementally shifted upwards for clarity. The direction of the
asymmetric tail changes between $B=0.9142$ and 0.9160~T, and the
symmetric shape appears at 0.9151~T. The results of the fitting with
Fano's line shape with real $q$'s are superposed.}
\label{FanoAB2D}
\end{figure}

The result of the magnetic controlling of the Fano effect was already
reported (Fig.~4 of Ref.~\onlinecite{KobayashiPRL2002}): Fano's
asymmetric Coulomb oscillation appears in sweeping $V_g$ at a fixed
magnetic field, while the field sweeping at a fixed $V_g$ results in
the AB oscillation.  Figure~\ref{FanoAB2D} shows the asymmetric
oscillation obtained at $B=0.9142$, 0.9151, and 0.9160~T. Defining the
first as $\Delta \theta = 0$, the second and the third correspond to
$\Delta \theta = \pi/2$ and $\pi$, respectively, as is calculated from
the AB period $\sim 3.6$~mT. The peaks at $\Delta \theta = 0$ and
$\pi$ are very asymmetric while those at $\Delta \theta = \pi/2$ are
almost symmetric.  The peaks at $V_g = -0.058$~V are fitted to
Eq.~(\ref{CondFanoEqn}) as indicated by the dashed curves.  The
obtained $q$'s are $-2.6$, $-18$, and $2.0$ for $\Delta \theta = 0$,
$\pi/2$, and $\pi$, respectively. This result is sufficient for one to
cast doubts on the applicability of Eq.~(\ref{CondFanoEqn2}) with a
real $q$. Since $q$ serves as a measure for the strength of the
configuration interaction, it is unphysical that $q$ changes
drastically and periodically by the small variation of the magnetic
field.  Treating $q$ as a real number also conflicts the observation
that the peaks become symmetric at specific fields, because
Eq.~(\ref{CondFanoEqn2}) never gives a symmetric line shape for a real
$q$ unless $|q| \rightarrow \infty$ or $q=0$ and the divergence of $q$
should be ruled out, which means that there is no configuration
interaction.

To overcome these unphysical situations, we should take it into
account that the field only affects the phase of the electron and
never changes the strength of the configuration interaction.  In the
original situation of the Fano theory,\cite{FanoPR1961} this does not
occur because the scattering center and the continuum are spatially
overlapping. Hence $q$ can be treated as a real number.  Even in the
framework of the Hamiltonian expressed in Eq.~(\ref{FanoEq1}), this
can be performed by treating $V_E$ as a complex number and by assuming
that the field only modifies $\arg(V_E)$ but not $|V_E|$.  As a
consequence, the generalized Fano formula is proposed in the following
expression as an expansion of Eq.~(\ref{CondFanoEqn2}),
\begin{equation}
G_{Fano} (V_g, B) = A \frac{|\epsilon+q|^2}{{\epsilon}^2+1} =
A \frac{(\epsilon+\mbox{Re}q)^2+(\mbox{Im}q)^2}{{\epsilon}^2+1}.
\label{FanoComplex}
\end{equation}
Qualitatively, even when the coupling strength $|q|$ is almost
independent of $B$, the $B$-dependence of $\arg(q)$ that comes from
$V_E$ in Eq.~\ref{defq} yields asymmetric and symmetric line shapes of
$G$ for $|\mbox{Re}q| \gg |\mbox{Im}q|$ and $|\mbox{Re}q| \ll
|\mbox{Im}q|$, respectively.  We treated $q$ as real in the analysis
of the peaks in Fig.~\ref{TypicalFano}, which is justified as their
asymmetry is large enough.

We note that $q$ in Eq.~(\ref{defq}) is not necessarily real.
Conventionally $q$ has been considered to be real, which is valid only
when the system has the time-reversal symmetry (TRS) and thus the
matrix elements defining $q$ in Eq.~(\ref{defq}) can be taken as real.
This is the case in many experimental situations of microscopic Fano
systems as far as their ground state has TRS, because it needs
enormous magnetic field to add considerable AB phase during the
transition between the continuum and the discrete states. Hence the
breaking of TRS becomes important with enlargement of systems.  The
claim that Fano's asymmetric parameter $q$ should be complex was
theoretically considered in
spectroscopy.\cite{AgarwalPRA1984,AbstreiterLSS,JinPRB1994,IhraPRB2002}
Also in the electron transport through mesoscopic systems, such
generalization has been discussed or
alluded.\cite{RacecPRB2001,WohlmanCM2001,ClerkPRL2001,%
HofstetterPRL2001,KimPRB2002,UedaJPSJ2003,KangCM2002}  To the best of
our knowledge, though, our result is the first convincing experimental
indication that $q$ should be a complex number, when TRS is broken
under the magnetic field.

\begin{figure}[htb]
\center
\includegraphics[width=0.90\linewidth]{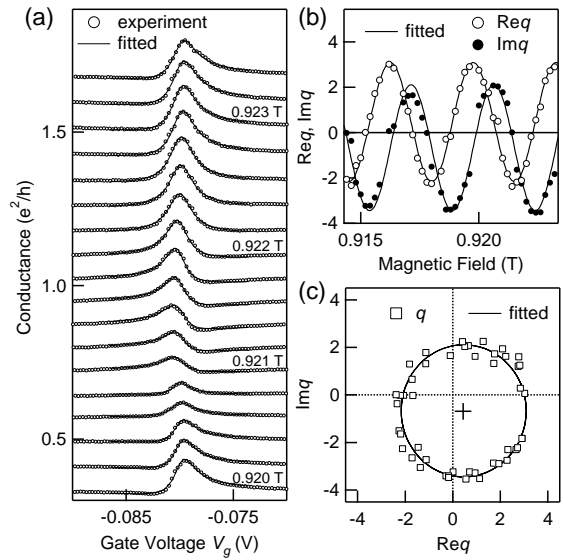}
\caption{(a) Conductance of the system measured at 30~mK at different
magnetic fields that cover one AB period.  The open circles and the
solid curves are the experiments and the results of the fitting,
respectively.  They are incrementally shifted upwards for clarity. (b)
Obtained $\mbox{Re}q$ and $\mbox{Im}q$ are plotted. The solid curves
are the fitted sinusoidal curves.  (c) Result of (b) plotted in the
complex $q$ plane by treating $B$ as an internal parameter. The cross
indicates the ellipse center of the complex $q$.}
\label{FanoSim}
\end{figure}

In order to be more quantitative, we carried out numerical fitting
with complex $q$'s to the line shape of the peaks occurring in the
range $B = 0.91$--0.93~T. The fitting function is
Eq.~(\ref{CondFanoEqn}) with Eq.~(\ref{FanoComplex}) for
$G_{Fano}(V_g)$, with $\mbox{Re}q$, $\mbox{Im}q$, $V_0$, ${\it
\Gamma}$, $A$, and $G_{bg}$ as free parameters. The values of $V_0$
and ${\it \Gamma}$ can be determined straightforwardly. As for the
other four parameters, $\mbox{Re}q$, $\mbox{Im}q$, $A$, and $G_{bg}$,
the numerical fitting alone does not allow unique determination of all
of them, because the functional form of Eq.~(\ref{CondFanoEqn})
essentially contains only three parameters besides $V_0$ and ${\it
\Gamma}$. We took the following fitting procedure. As has been argued
in the discussion leading to Eq.~(\ref{FanoComplex}), the amplitude
$A$ of the Fano resonance is expected to be nearly independent of
$B$. Therefore, it is reasonable to assume that $A$ takes a common
value for all the curves to be fitted. The value of $A$ was determined
from the fitting of the most asymmetric case (since $q$ can be taken
as real) as $0.011$$e^2/h$, and this value was used for other
curves. The resulting fitting is reasonably good over the whole range
as shown by solid curves covering one AB period in
Fig.~\ref{FanoSim}(a).  The $V_0$ and ${\it \Gamma}$ are nearly
independent of $B$, while the $G_{bg}$ is found to change slightly
with $B$ due to the long tails of the neighboring Fano peaks.

In Fig.~\ref{FanoSim}(b), the obtained $\mbox{Re} q$ and $\mbox{Im}
q$ are plotted against $B$. Both of them well depend on $B$
sinusoidally, where the phase difference between the two sinusoids is
very close to $\pi/2$. Indeed, the solid curves superposed on them are
the result of the fitting by a sinusoidal function, that is,
$\mbox{Re} q (\Delta \theta)= 0.4-2.6\cos (\Delta \theta)$ and
$\mbox{Im} q (\Delta \theta) = -0.7-2.7\sin (\Delta \theta)$.  As
defined above, $\Delta \theta$ is the phase difference picked up from
the magnetic field based on the measured AB period, while $\Delta
\theta=0$ corresponds to $B=0.9145$~T for these sinusoids.

The result is summarized as the plot of $q$ in the complex plane shown
in Fig.~\ref{FanoSim}(c). The solid curve, which is elliptic (almost
circular), is the result of the fitting in Fig.~\ref{FanoSim}(b). As
indicated by the cross, the ellipse center is slightly shifted from
the origin $q=0$. The same fitting procedure for the other peaks at
$V_g = -0.058$~V and $-0.038$~V in Fig.~\ref{FanoAB2D} yields complex
$q$ which also traces an ellipse with the center slightly shifted from
the origin.

The generalized Fano formula for the two-terminal conductance of the
QD-AB-ring system was theoretically proposed to be the same functional
form with
Eq.~(\ref{FanoComplex}).\cite{HofstetterPRL2001,UedaJPSJ2003}  In the
formula, the effect of the magnetic field is included in the phase
factor $\Delta \theta$ in the complex $q$ that is given as $q (\Delta
\theta) = q_1 \cos (\Delta \theta) + i q_2 \sin (\Delta \theta)$.
$q_1$, $q_2$ and $A$, which are real parameters independent of $B$,
are defined by both the tunneling coupling between the QD and its
leads and the coupling strength between the QD and the continuum
energy state. The theoretical $q$ traces an ellipse in the complex
plane being consistent with our observation, while the theoretical
ellipse center is at the origin.  It is emphasized that this
difference between the experiment and the theory is not caused by our
fitting procedure. As is clear from Fig.~\ref{FanoAB2D}, the heights
of the peak at $V_g = -0.079$~V are different when the phase
difference changes by $\pi$.  This is also the case for the $V_g =
-0.038$~V peak.  These observations are never reproduced by using the
theoretical form of $q (\Delta \theta)$, because, for an arbitrary
$\Delta \theta$, the theoretical line shape for $q (\Delta \theta)$ is
a mirror image of that for $q (\Delta \theta+\pi)$ with respect to
$\epsilon=0$.  Thus, the difference is due to the effect that is not
included in the theoretical model.  The influence of the multichannel
transport and/or the orbital phase~\cite{KubalaPRB2003} might be a
possible candidate. The incoherent transmission through the system may
be also responsible for that, because $q$ generally becomes complex
due to the decoherence even under TRS.\cite{ClerkPRL2001}  As the
present system is highly, but not perfectly, coherent, this would
shift the ellipse center of the complex $q$ from the origin.

\subsection{Electrostatic control of the Fano interference}
\label{subsecC}
Next, we focus on the electrostatic control of the Fano effect by the
control gate voltage $V_C$.  The variation of the electrostatic
potential energy by $V_C$ results in change in the kinetic energy of
electrons, and hence their wave number which traverses the region
underneath the gate electrode. As this gives rise to the phase
difference between the path through the QD and the one through the
arm,~\cite{kats1992} the electrostatic controlling of the Fano effect
is expected.

\begin{figure}[!htb]
\center \includegraphics[width=0.90\linewidth]{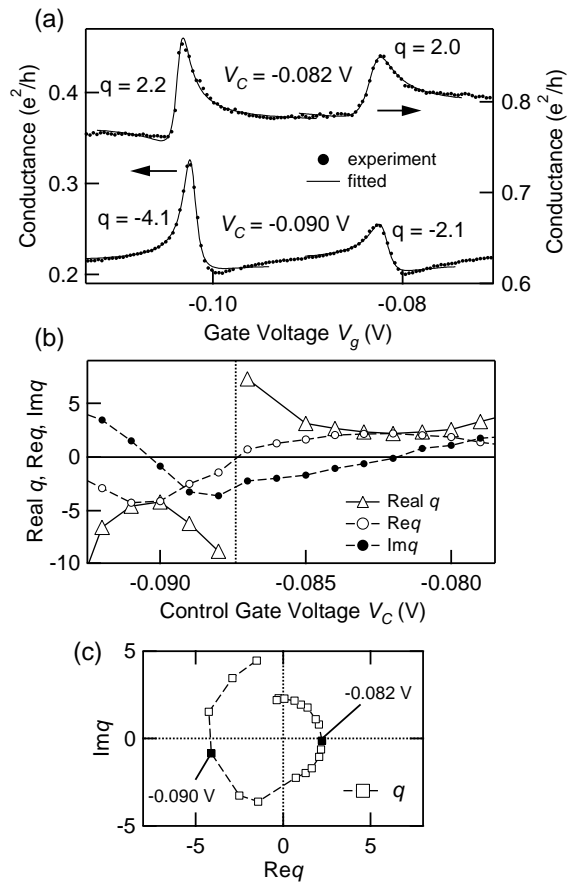}
\caption{ (a) Coulomb oscillations at $V_C=-0.082$~V and $-0.090$~V at
30~mK and $B=0.91$~T. The results of the fitting to
Eq.~(\ref{CondFanoEqn}) are shown in the solid lines with the obtained
$q$ values.  (b) Real $q$ for the $V_g = -0.102$~V peak is plotted by
the triangle as a function of $V_C$. The sign of $q$ changes between
$V_C = -0.088$~V and $-0.087$~V as indicated by the vertical dashed
line. The obtained $\mbox{Re}q$ and $\mbox{Im}q$ for the same peak are
superposed with white and black circles, respectively.  (c) Result of
(b) plotted in the complex $q$ plane by treating $V_C$ as an internal
parameter. $q$'s for $V_C = -0.082$~V and $-0.090$~V are colored
black. }
\label{FigEleControl}
\end{figure}

Figure~\ref{FigEleControl} (a) shows the asymmetric Coulomb
oscillation at $V_C = -0.082$~V and $-0.090$~V.  The direction of
asymmetric tail of the Fano effect is opposite between the two.  The
result of the fitting to Eq.~(\ref{CondFanoEqn}) with real $q$ for
the peak at $V_g = -0.102$~V is shown by the solid curve.  In
Fig.~\ref{FigEleControl} (b), the real $q$ obtained for various $V_C$
is plotted by the triangles.  The line shape becomes symmetric with
large $|q|$ at $V_C = -0.088$~V and $-0.087$~V, between which the sign
of $q$ changes.

The above behavior is similar to the one in the magnetic controlling.
However, we note that there exists an essential difference between the
magnetic controlling and the electrostatic one. While in the former a
complex $q$ is a natural consequence of the broken TRS, this is not
the case in the latter because the electrostatic potential modulation
itself never breaks TRS when the magnetic flux piercing the AB ring is
integer times the flux quantum $h/e$.  Nevertheless, when the flux is
fixed to a noninteger value of flux quanta, a symmetric Lorentzian
line shape can appear at specific $V_C$'s and thus $q$ should be
complex.  This can be proven in a model calculation by combining the
Breit-Wigner formula with the conductance formula for the two-terminal
AB ring~\cite{GefenPRL1984} as performed in
Ref.~[\onlinecite{WohlmanCM2001}].  Finite decoherence is also
responsible for the occurrence of the symmetric
line shape~\cite{ClerkPRL2001} as already discussed.

Figure~\ref{FigEleControl}(b) shows the result of the complex $q$ for
the various $V_C$'s obtained by the similar fitting procedure as done
in the magnetic controlling. As expected, both $\mbox{Re}q$ and
$\mbox{Im}q$ continuously change in sweeping $V_C$.  $\mbox{Re}q$
crosses $0$ with finite $\mbox{Im}q$ between $V_C = -0.088$~V and
$-0.087$~V while $|\mbox{Re}q|$ takes its local maximum with
$\mbox{Im}q \sim 0$ around $V_C = -0.082$~V and $-0.090$~V, where the
line shape is the most asymmetric. So, the electrostatic controlling
of the Fano effect can be characterized by the complex $q$ as we have
seen for the magnetic one. 
As we have discussed in Fig.~\ref{FanoSim}(c), the result is
summarized in Fig.~\ref{FigEleControl}(c) as the plot of $q$ in the
complex plane.  Again, the complex $q$ encircles around the origin of
the complex plane.

Let us estimate the order of magnitude of the phase difference
introduced by the change of $V_C$.  According to a simple capacitance
model between the gate and the
2DEG,\cite{YacobyPRL1991,KobayashiJPSJ2002} this effect yields the
phase difference $\Delta\theta$;
\begin{equation}
\Delta\theta(V_C) = 2 \pi \frac{W}{\lambda_F} \left(1 -
\sqrt{1-\frac{V_C}{V_{dep}}}\right),
\label{EleStVdep}
\end{equation}
where $W$ and $V_{dep}$ are the width of the gate and the pinch-off
voltage of the corresponding conduction channel ($-0.10$~ V for the
last channel in the present case), respectively. $\lambda_F \sim
40$~nm is the Fermi wavelength of the present 2DEG.  Now, if we take
the effective width $W \sim 120$~nm, $|\Delta\theta (V_C = -0.082)-
\Delta\theta (V_C = -0.090)| = 0.65 \pi$ for the last and probably the
most effective channel.  Although this is expected to be $\pi$ for the
ideal case
as seen from Fig.~\ref{FigEleControl}(c), 
the variation of the Fano line shape in Fig.~\ref{FigEleControl}(a) is
attributed as due to such large phase difference electrostatically
introduced by $V_C$.

We have seen that the electrostatic controlling of the Fano effect is
possible.  It is pointed out that this type of phase measurement is
very difficult in ordinary oscillation of conductance.\cite{kats1992}
A phase shift by $\pi$ due to potential scattering means that the
Fermi energy has passed a resonant energy level, which causes steep
variation of the conductance by order of $2e^2/h$.  Hence the
oscillation due to the interference is almost swamped by in the large
variation of the conductance.  Indeed in the case of
Fig.~\ref{FigEleControl}, the increase of the baseline for $\pi$ phase
shift is about 0.6$e^2/h$ while the Fano asymmetry due to the
interference is only about 0.02$e^2/h$.  Furthermore, $V_{dep}$ in
Eq.~\eqref{EleStVdep} is different for each conductance channel,
meaning that the phase shift caused electrostatically is inevitably
channel dependent.  Hence the total conductance for the multichannel
case is a superposition of many oscillations with various phase
shifts.

The clear advantage of the present electrostatic tuning in overcoming
the severe ``signal-to-noise-ratio" lies in the peculiar Fano
line shape and in the insensitivity of the arm conductance to $V_g$.
More importantly, the QD acts as a filter for the conductance
channels, which means that in the Fano effect we selectively observe
the interference through the single channel passing through the QD.
The present result indicates that the analysis of the Fano resonance
with Eq.~\eqref{FanoComplex} provides a sensitive and noise-resistant
method to detect quantum-mechanical phase of electrons.

\subsection{Phase evolution over multiple resonances}
\label{subsecD}
Thus far we have mainly focused on the transport in the small areas of
the gate voltage around respective peaks. Here, we discuss the phase
evolution of electrons between neighboring peaks.  In
Ref.~\onlinecite{KobayashiPRL2002}, it has been reported that the
phase changes by $\pi$ rapidly but continuously across the resonance
and that all the adjacent Fano resonances are in phase, as the phase
makes another slow change by $\pi$ between the valley.  In contrast to
the experiment for the phase measurement of the
QD,\cite{YacobyPRL1995} the present data not only reflect the phase
evolution of electrons through a QD, but also represent the response
to the magnetic field of the whole QD-AB-ring system. In that case, a
continuous change by $\pi$ at the resonance is
likely.\cite{UedaJPSJ2003} What still remains surprising is the
another gradual phase change by $\pi$ at each valley.\cite{comment1}
This phenomenon is in clear contrast to the reported phase lapse in
the middle of the valley,\cite{SchusterNature1997} which has not been
perfectly understood yet despite extensive
studies.\cite{YeyatiPRB1995,BruderPRL1996,OregPRB1997,WuPRL1998,%
HackenbroichEPL1997,LeePRL1999,TaniguchiPRB1999} Nevertheless, in most
cases, the adjacent Coulomb peaks are in phase. This is also
consistent with our observation that the direction of the Fano
asymmetry is usually the same in each curve as clearly seen in
Figs.~\ref{TypicalFano}, \ref{FanoMag}(b), ~\ref{FanoAB2D}, and
\ref{FigEleControl}(a).

It should be added that in rare cases the adjacent peaks are in
antiphase, as seen in the Fano peaks 8 and 9 in
Fig.~\ref{TypicalFano}.  This is the observation that has never
been known before.\cite{YacobyPRL1995,SchusterNature1997}  While the
necessary condition for it remains to be clarified, this should be a
new clue to the phase lapse problem.

\subsection{Temperature dependence of the AB effect}
\label{subsecE}
\begin{figure}[htb]
\center \includegraphics[width=0.90\linewidth]{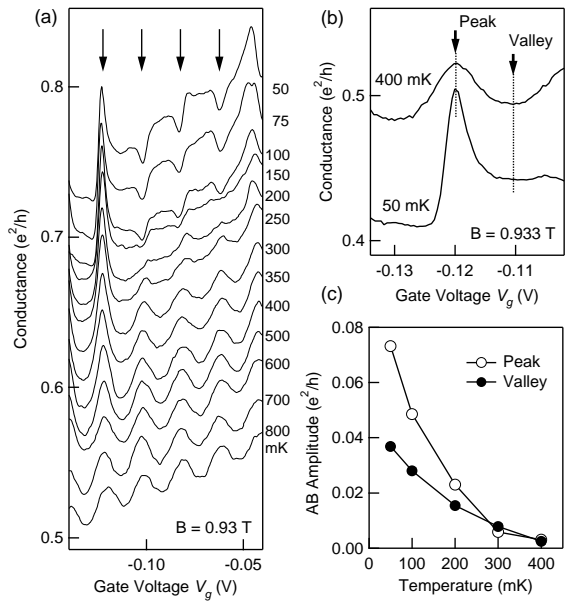}
\caption{(a) Conductance of the system measured between 50~mK and
800~mK at $B =0.93$~T.  The data at $T \leq 700$ mK are incrementally
shifted upward for clarity. The Fano features marked by the arrows at
50~mK gradually disappear as the temperature increases.  (b) Fano
structure at 50~mK and 400~mK.  (c) Temperature dependence of the AB
amplitude measured at the peak and the valley indicated in (b).}
\label{FanoTdep} \end{figure}

The phase coherence over the system is essential in the controlling of
the Fano interference discussed above.  Figure~\ref{FanoTdep}(a)
shows the temperature dependence of the Fano effect between 50~mK and
800~mK.  While clear Fano features appear at 50~mK as marked by the
arrows, they diminish rapidly as the temperature
increases. Especially, the asymmetric dips at 50~mK evolves into peaks
at $T \geq 200$~mK.  While the thermal broadening of the Fano
line shape is not negligible at temperatures higher than ${\it
\Gamma}$,\cite{ZachariaPRB2001} it cannot explain such drastic
temperature dependence.  The main cause for this phenomenon is
the decoherence induced by increasing the temperature, resulting in the
destruction of the Fano state. This is confirmed by the temperature
dependence of the AB effect.  In Fig.~\ref{FanoTdep}(c), we plotted
the AB amplitude at 50, 100, 200, 300, and 400~mK for the peak and the
valley indicated in Fig.~\ref{FanoTdep}(b).  At 50~mK, the AB
amplitude is found to be comparable to the net peak height, indicating
that the electronic states are highly coherent over the ring.  At $T>
400$~mK, the AB amplitude was hardly visible. This is consistent with
the observation in Fig.~\ref{FanoTdep}(a).

The AB amplitude monotonically decreases both at the peak and valley
as shown in Fig.~\ref{FanoTdep}(c). Interestingly, it decays more
slowly at the valley than at the peak. While the mechanism for such a
temperature dependence is yet unclear, a logarithmic $T$ dependence
that might be reminiscent of the Kondo effect was reported for the
Fano effect in a single QD.\cite{GoresPRB2000} On the other hand, the
exponential-decay function of $T$ was reported for the quasiballistic
AB
ring.\cite{HansenPRB2001,KobayashiJPSJ2002,SeeligPRB2001,SeeligCM2003}
However, the observed temperature range Fig.~\ref{FanoTdep}(c) is not
wide enough to distinguish these two dependencies.

\section{Conclusion}
In summary, we have studied mesoscopic Fano effect observed in a
QD-AB-ring system, especially in terms of the magnetic and the
electrostatic tuning.  In the conventional analysis with real $q$, the
correlation between the original Coulomb peak heights and $|q|$ is
observed.  The magnetic-field tuning has revealed that the effect of
TRS breaking can be taken into account by extending the Fano parameter
$q$ to a complex number.  We have shown that the experimental curves
can actually be fitted by adopting a complex $q$,\cite{KimPRL2003}
which shows slight systematic deviation from theoretical prediction.
We have applied thus established method to the electrostatic tuning of
Fano effect.  This method clearly describes the electrostatic tuning
of electron phase, which demonstrates that the Fano effect can be a
powerful tool for phase measurement in mesoscopic circuits.  Finally,
we have reported the difference in the temperature dependence of the
AB effect between the Coulomb peak and the Coulomb valley.  Thus, we
have clarified several aspects of the Fano effect observed in
transport, while some problems remain to be solved in future such as
the phase evolution over the multiple resonances.

It is interesting that the Fano effect first established in atomic
physics is now observed in an artificial atom system. Such a
mesoscopic analog will bring us renewed understandings.  For example,
the complex $q$ should also be applied to other experiments where TRS
is broken.

\section*{ACKNOWLEDGMENTS}
We thank A. Aharony, M. B{\"u}ttiker, O. Entin-Wohlman, M. Eto,
W. Hofstetter, T.-S. Kim, J. K{\"o}nig, and T. Nakanishi for helpful
discussion.  This work was supported by a Grant-in-Aid for Scientific
Research and by a Grant-in-Aid for COE Research (``Quantum Dot and Its
Application'') from the Ministry of Education, Culture, Sports,
Science, and Technology of Japan. K.K. was supported by a Grant-in-Aid
for Young Scientists (B) (No.~14740186) from Japan Society for the
Promotion of Science.


\begin{thebibliography}{99}
\bibitem{comment1} Since this feature is not included in
Eq.~(\ref{FanoComplex}), the fitting in Fig.~\ref{FanoSim}(a) should
be performed to the limited range of $V_g$ around a single resonance,
while the results of the fitting such as the behavior of the complex
$q$ are not greatly affected by a slight change of the fitting range.
\end{thebibliography}
\end{document}